\documentclass[doublecol]{epl2}
\usepackage{graphicx,amsmath,amssymb,amsfonts,latexsym,color,dcolumn,bm,epsfig,subfigure}
\usepackage[latin1]{inputenc}

\def\bbm[#1]{\mbox{\boldmath $#1$}}
\newcommand{\ket}[1]{\displaystyle{|#1\rangle}}
\newcommand{\bra}[1]{\displaystyle{\langle #1|}}
\newcommand{\Tr}{\text{Tr}}
\newcommand{\Rea}{\text{Re}}
\newcommand{\Ima}{\text{Im}}
\newcommand{\TE}{\text{TE}}
\newcommand{\TM}{\text{TM}}

\title{Casimir-Lifshitz force out of thermal equilibrium and heat transfer between arbitrary bodies}
\shorttitle{Casimir-Lifshitz force out of thermal equilibrium and heat transfer between arbitrary bodies} 

\author{R. Messina\inst{1} \and M. Antezza\inst{2,3}}
\shortauthor{R. Messina \etal}

\institute{
  \inst{1} SYRTE - Observatoire de Paris, 61, avenue de l'Observatoire, F-75014 Paris, France\\
  \inst{2} Universit\'e  Montpellier 2, Laboratoire Charles Coulomb UMR 5221, F-34095, Montpellier, France\\
  \inst{3} CNRS, Laboratoire Charles Coulomb UMR 5221, F-34095, Montpellier, France}


\abstract{We study the Casimir-Lifshitz force and the radiative heat transfer occurring between two arbitrary bodies, each one
held at a given temperature, surrounded by environmental radiation at a third temperature. The system, in stationary configuration
out of thermal equilibrium, is characterized by a force and a heat transfer depending on the three temperatures, and explicitly
expressed in terms of the scattering operators of each body. We find a closed-form analytic expression valid for bodies of any geometry and dielectric
properties. As an example, the force between two parallel slabs of finite thickness is
calculated, showing the importance of the environmental temperature as well as the occurrence of a repulsive interaction. An
analytic expression is also provided for the force acting on an
atom in front of a slab. Our predictions can be relevant for experimental and technological
purposes.}

\begin{document}

\maketitle

\section{Introduction}The quantum and thermal fluctuations of the
electromagnetic field result in a force between any couple of
polarizable bodies. This dispersion effect, originally predicted
by Casimir for two parallel perfectly conducting plates at zero
temperature, and later generalized by Lifshitz and coworkers to
real dielectric materials at finite temperatures
\cite{Lifshitz5561}, becomes relevant for bodies separated by less
than few microns. Today, the Casimir-Lifshitz force plays a major
role in all fundamental and technological issues occurring at
these separations, becoming widely interesting, from biological
systems to microelectromechanics \cite{MEMS1,MEMS2}.

The force shows two components, related to the purely quantum and
thermal field fluctuations, respectively. The former dominate at
short distances, and have been measured in several configurations
since early times. The latter are relevant at large separations,
with a weak total effect  measured only recently
\cite{ObrechtPRL07,LamoreauxNaturePhys11}. The Casimir-Lifshitz
force, largely studied for systems at thermal equilibrium, has
been considered also for systems out of thermal equilibrium
\cite{Rytov89,Landau80}, as done for the
atom-atom, plane-plane and atom-surface force in
\cite{RosenkranzJCP68,DorofeyevJPhysA98,HenkelJOptA02}. More
recently, other particular configurations have been studied out of
thermal equilibrium, involving infinitely thick planar and non
planar bodies, and atoms
\cite{AntezzaPRLA06081,AntezzaPRLA06082,BimontePRA09,BuhSher08091,BuhSher08092}.
Indeed, non-equilibrium effects received a renewed interest since
the recent discovery that systems driven out of thermal
equilibrium may show new qualitative and quantitative behaviors,
namely the possibility of a repulsive force, and of a strong force
tunability \cite{AntezzaPRL05}. Such peculiar characteristics
allowed a non-equilibrium system to be used for the first
measurement of thermal effects, by exploiting trapped ultracold
atomic gases \cite{ObrechtPRL07}.

Such a force, related to the correlations of the electromagnetic
field,  shares a common formalism with radiative heat transfer
\cite{PolderPRB71}, recently object of several investigation for
both fundamental and technological issues
\cite{RousseauNaturePhoton09}.

Due to a less direct formalism, calculations of forces and heat
transfer for systems out of equilibrium have been performed only
for some ideal configurations (atoms at zero temperature,
infinitely extended surfaces, \dots), and a general theory able to
take into account arbitrary materials and geometries at different
temperatures in an arbitrary thermal environment, is still
missing. In this paper we derive such a theory, and we obtain a
closed-form explicit expression for the most general problem of
the force and heat transfer between two arbitrary bodies at two
different temperatures placed in a thermal environment having a
third (in general different) temperature. In addressing this
problem we describe the two bodies by means of their scattering
operators. This approach, successfully used to calculate the
Casimir-Lifshitz force both at \cite{ScatteringReynaud,RahiPRD09}
and out \cite{BimontePRA09} of thermal equilibrium, even if
formally equivalent to a Green function technique, presents the
advantage of requiring only single-body operators rather than the
complete solution for the composite system.

\section{The physical system}
\begin{figure}\begin{center}\scalebox{0.4}{\includegraphics{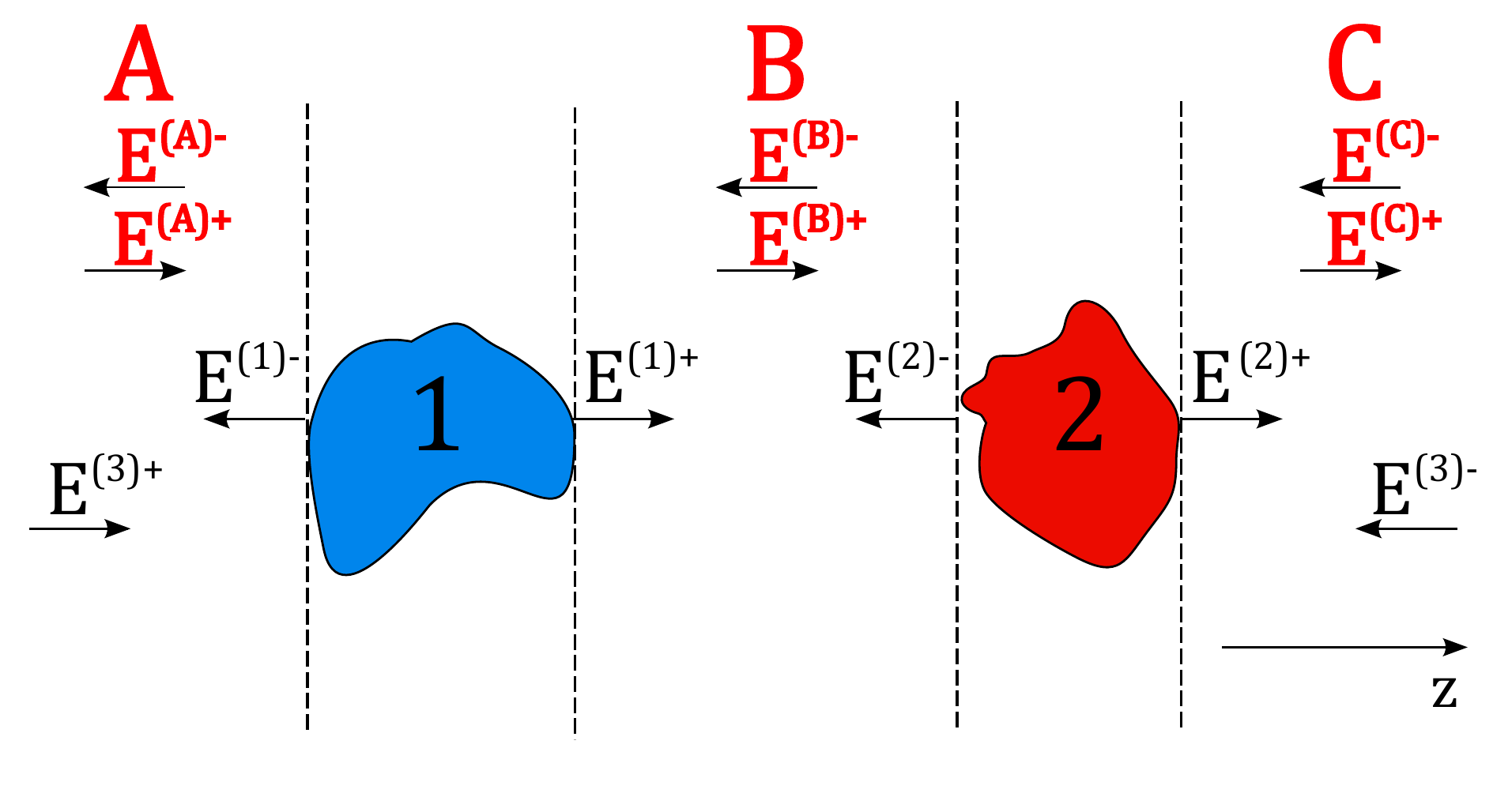}}\vspace{-10pt}
\caption{(color online) Scheme of the system. The body 1 (2) is
held at temperature $T_1$ ($T_2$), while the environment has
temperature $T_3$.}\label{FigGeometry}\end{center}\end{figure}

The system is made of two bodies,
labeled with indexes $1$ and $2$, which we assume separated by at
least an infinite plane. This assumption, introduced to exclude
two concatenated bodies, is not necessary in approaches based on
scattering theory \cite{RahiPRD09}, nonetheless it makes the
formalism much easier to follow and it is verified practically in
all the common experimental configurations (e.g. plane-plane,
sphere-plane, atom-surface). We are going to calculate the
component of the forces acting on the two bodies along the axis
perpendicular to such a plane, to which we refer as the $z$ axis,
as well as the heat flux on each of them. A scheme of the system
is shown in figure \ref{FigGeometry}, where three distinct regions
A, B and C are defined.
 The
body $1$ ($2$) is assumed to be at a $local$ thermal equilibrium
at temperature $T_1$ ($T_2$), and on the two bodies impinges a
thermal radiation at the third temperature $T_3$. One can imagine
a system where the two bodies are inside, and far from the
surfaces, of a much bigger cell (the reservoir), held locally at
thermal equilibrium at temperature $T_3$. The three temperatures
remain constant in time, so that the system assumes a stationary
regime. If at least one of the two bodies is
microscopic, the validity of his hypothesis requires a more
careful analysis: generally speaking it remains acceptable for
time scales much shorter that the typical time of evolution of the
microscopic body.

The force $\mathbf{F}$ acting on any of the two bodies and the
heat absorbed by it per unit of time can be evaluated calculating
the fluxes
\begin{equation}\label{Fms}\mathbf{F}=\int_{\Sigma}\langle\mathbb{T}(\mathbf{R},t)\rangle_\text{sym}\,\cdot d\bbm[\Sigma],\hspace{.5cm}H=-\int_{\Sigma}\langle\mathbf{S}(\mathbf{R},t)\rangle_\text{sym}\,\cdot d\bbm[\Sigma]\end{equation}
of the symmetrized averages $\langle AB\rangle_\text{sym}=\langle
AB+BA\rangle/2$ of the Maxwell stress tensor $\mathbb{T}$ and
Poynting vector $\mathbf{S}$ in vacuum
\begin{equation}\label{TMax}\begin{split}T_{ij}(\mathbf{R},t)&=\epsilon_0\Bigl[E_iE_j+c^2B_iB_j-\frac{1}{2}\Bigl(E^2+c^2B^2\Bigr)\delta_{ij}\Bigr]\\
\mathbf{S}(\mathbf{R},t)&=\epsilon_0c^2\mathbf{E}\times\mathbf{B}\end{split}\end{equation}
($i,j=x,y,z$), through an arbitrary closed and oriented surface
$\Sigma$ enclosing the body. Choosing $\Sigma$ as a parallelepiped
having two of its faces orthogonal to the $z$ axis (and on two
opposite sides of a given body) and letting the surface of these
sides tend to infinity, one finds that the $m$ component
($m=x,y,z$) $F_m$ of the force (the heat transfer) has non
negligible contribution only from the difference of the fluxes of
the $\langle T_{mz}\rangle$ component of the tensor (the $\langle
S_z\rangle$ component of the Poynting vector) on the two sides of
the considered body. For the electric and magnetic
fields appearing in eq. \eqref{TMax} we use the following mode
decomposition
\begin{equation}\mathbf{E}(\mathbf{R},t)=2\textrm{Re}\Bigl[\int_0^{+\infty}\frac{d\omega}{2\pi}e^{-i\omega t}\mathbf{E}(\mathbf{R},\omega)\Bigr]\end{equation}
where
\begin{equation}\label{DefE}\mathbf{E}(\mathbf{R},\omega)=\sum_{p,\phi}\int\frac{d^2\mathbf{k}}{(2\pi)^2}e^{i\mathbf{K}^\phi\cdot\mathbf{R}}\hat{\bbm[\epsilon]}_p^\phi(\mathbf{k},\omega)E_p^\phi(\mathbf{k},\omega).\end{equation}
In this representation, known as angular spectrum representation
\cite{NietoVesperinas91,SipeJOptSocAmB87}, a mode has amplitude
$E_p^\phi(\mathbf{k},\omega)$, corresponding to the frequency
$\omega$, the transverse wavevector $\mathbf{k}=(k_x,k_y)$, the
transverse polarization $p$ taking the values 1 (TE) and 2 (TM),
and the direction of propagation along the $z$ axis $\phi=\pm1$
(with shorthand notation $\phi=\pm$ in the polarization vectors
and field amplitudes). In this approach, which proves to be
convenient in our planar-like geometry,
$k_z=\sqrt{\omega^2/c^2-\mathbf{k}^2}$ is a dependent variable,
and the three-dimensional wavevector is noted as
$\mathbf{K}^\phi=(\mathbf{k},\phi k_z)$. We have also introduced
the polarization unit vectors, defined as
$\hat{\bbm[\epsilon]}_\TE^\phi(\mathbf{k},\omega)=\hat{\mathbf{z}}\times\hat{\mathbf{k}}$
and
$\hat{\bbm[\epsilon]}_\TM^\phi(\mathbf{k},\omega)=\hat{\bbm[\epsilon]}_\TE^\phi(\mathbf{k},\omega)\times\hat{\mathbf{K}}^\phi$
where $\hat{\mathbf{z}}=(0,0,1)$ and
$\hat{\mathbf{A}}=\mathbf{A}/A$. The analogous expression for the
magnetic field can be directly deduced from Maxwell's equations.
The field $E^{(\gamma)\phi}$ in fig. \ref{FigGeometry} is the part
of the field emitted by the body $\gamma$ (for $\gamma=1,2$), of
the environmental field (for $\gamma=3$), or of the \emph{total}
field in each region (for $\gamma=A,B,C$) propagating in direction
$\phi$. The fluxes of $\langle T_{mz}\rangle$ and $\langle
S_z\rangle$ through a given plane $z=\bar{z}$ can now
be explicitly expressed in the following way
\begin{equation}\label{FluxTmz}\begin{split}&\Phi_m(\bar{z})=\int_{z=\bar{z}}d^2\mathbf{r}\,\langle T_{mz}\rangle_\text{sym}\\
&=-\sum_p\int\frac{d^2\mathbf{k}}{(2\pi)^2}\Biggl(\sum_{\phi=\phi'}\int_{ck}^{+\infty}\frac{d\omega}{2\pi}+\sum_{\phi\neq\phi'}\int_0^{ck}\frac{d\omega}{2\pi}\Biggr)\\
&\,\times\frac{2\epsilon_0c^2k_z}{\omega^2}\bra{p,\mathbf{k}}C^{(\gamma)\phi\phi'}\ket{p,\mathbf{k}}\times\begin{cases}\phi
k_m & m=x,y\\k_z & m=z\end{cases}\end{split}\end{equation}
\begin{equation}\label{FluxSz}\begin{split}&\varphi(\bar{z})=\int_{z=\bar{z}}d^2\mathbf{r}\,\langle S_z\rangle_\text{sym}\\
&=\sum_p\int\frac{d^2\mathbf{k}}{(2\pi)^2}\Biggl(\sum_{\phi=\phi'}\int_{ck}^{+\infty}\frac{d\omega}{2\pi}+\sum_{\phi\neq\phi'}\int_0^{ck}\frac{d\omega}{2\pi}\Biggr)\\
&\,\times\frac{2\epsilon_0c^2\phi
k_z}{\omega}\bra{p,\mathbf{k}}C^{(\gamma)\phi\phi'}\ket{p,\mathbf{k}}\end{split}\end{equation}
where $\gamma$ denotes the region in which $\bar{z}$ is
located. Each of these contributions do not depend on the
coordinate of the plane chosen to calculate the flux, provided
that this surface is located inside a given region $A$, $B$ or
$C$. These quantities are expressed as a function of the matrix
$C^{(\gamma)\phi\phi'}$
\begin{equation}\label{CommC}\begin{split}&\langle E^{(\gamma)\phi}_p(\mathbf{k},\omega)E^{(\gamma)\phi'\dag}_{p'}(\mathbf{k}',\omega')\rangle_\text{sym}\\
&=\frac{1}{2}\langle
E^{(\gamma)\phi}_p(\mathbf{k},\omega)E^{(\gamma)\phi'\dag}_{p'}(\mathbf{k}',\omega')+E^{(\gamma)\phi'\dag}_{p'}(\mathbf{k}',\omega')E^{(\gamma)\phi}_p(\mathbf{k},\omega)\rangle\\
&=2\pi\delta(\omega-\omega')\bra{p,\mathbf{k}}C^{(\gamma)\phi\phi'}\ket{p',\mathbf{k}'}\end{split}\end{equation}
being the correlating function between two amplitudes of the
\emph{total} field propagating in directions $\phi$ and $\phi'$,
and associated to a couple of modes $(\mathbf{k},p)$ and
$(\mathbf{k}',p')$. We have explicitly inserted the conservation
of frequency, direct consequence of the time invariance
(moving bodies are excluded from our calculation)
characterizing our system. In eq. \eqref{CommC} the correlators
are written as matrix elements of a matrix $C^{\phi\phi'}$: in our
notation, the matrices are defined on the space $(p,\mathbf{k})$
being $p=1,2$ and $\mathbf{k}\in\mathbb{R}^2$ and thus the product
of two matrices $A$ and $B$ is given by
\begin{equation}\begin{split}\bra{p,\mathbf{k}}\mathcal{A}\mathcal{B}\ket{p',\mathbf{k}'}&=\sum_{p''}\int\frac{d^2\mathbf{k}''}{(2\pi)^2}\bra{p,\mathbf{k}}\mathcal{A}\ket{p'',\mathbf{k}''}\\
&\,\times\bra{p'',\mathbf{k}''}\mathcal{B}\ket{p',\mathbf{k}'}.\end{split}\end{equation}

\section{At equilibrium}For a system at thermal equilibrium, the
correlators are directly given by the fluctuation-dissipation
theorem
\begin{equation}\label{FluDiss}
\begin{split}\langle E_i(\mathbf{R},\omega)E^{*}_j(\mathbf{R}',\omega')\rangle_\text{sym}&=2\pi\delta(\omega-\omega')\frac{2}{\omega}N(\omega,T)\\
&\,\times\Ima
G_{ij}(\mathbf{R},\mathbf{R}',\omega),\end{split}\end{equation}
where the purely quantum and the thermal fluctuations are clearly
distinguishable in the factor
$N(\omega,T)=\hbar\omega[1/2+n(\omega,T)]$, with
$n(\omega,T)=(\exp[\hbar\omega/(k_BT)]-1)^{-1}$, and $G_{ij}$ are
the components of the Green function $\bar{\mathbf{G}}$
associated to the system of the two bodies, solution of
the differential equation
\begin{equation}\Bigl[\nabla_\mathbf{R}\times\nabla_\mathbf{R}-\frac{\omega^2}{c^2}\epsilon(\omega,\mathbf{R})\Bigr]\bar{\mathbf{G}}(\mathbf{R},\mathbf{R}',\omega)=\frac{\omega^2}{\epsilon_0c^2}\,\bar{\mathbf{I}}\,\delta(\mathbf{R}-\mathbf{R}')\end{equation}
being $\bar{\mathbf{I}}$ the identity dyad and
$\epsilon(\omega,\mathbf{R})$ the dielectric function of the system. By expressing the Green function in terms of the
reflection scattering matrices operators $\mathcal{R}$ of the two
bodies, one recovers the force acting on the body 1 along the $z$
axis \cite{ScatteringReynaud,RahiPRD09}:
\begin{equation}\label{Feq}\begin{split}
F^{\text{(eq)}}&(T)=-2\Rea\Tr\Bigl\{\frac{k_z}{\omega}N(\omega,T)\\
&\,\times\Bigl[U^{(12)}\mathcal{R}^{(1)+}\mathcal{R}^{(2)-}+U^{(21)}\mathcal{R}^{(2)-}\mathcal{R}^{(1)+}\Bigr]\Bigr\},\end{split}\end{equation}
where $U^{(12)}=(1-\mathcal{R}^{(1)+}\mathcal{R}^{(2)-})^{-1}$ and
$U^{(21)}=(1-\mathcal{R}^{(2)-}\mathcal{R}^{(1)+})^{-1}$. The
definitions of the trace and the reflection operators in
\eqref{Feq} will become explicit in the following.

\section{Out of equilibrium}
\begin{figure}\begin{center}\scalebox{0.4}{\includegraphics{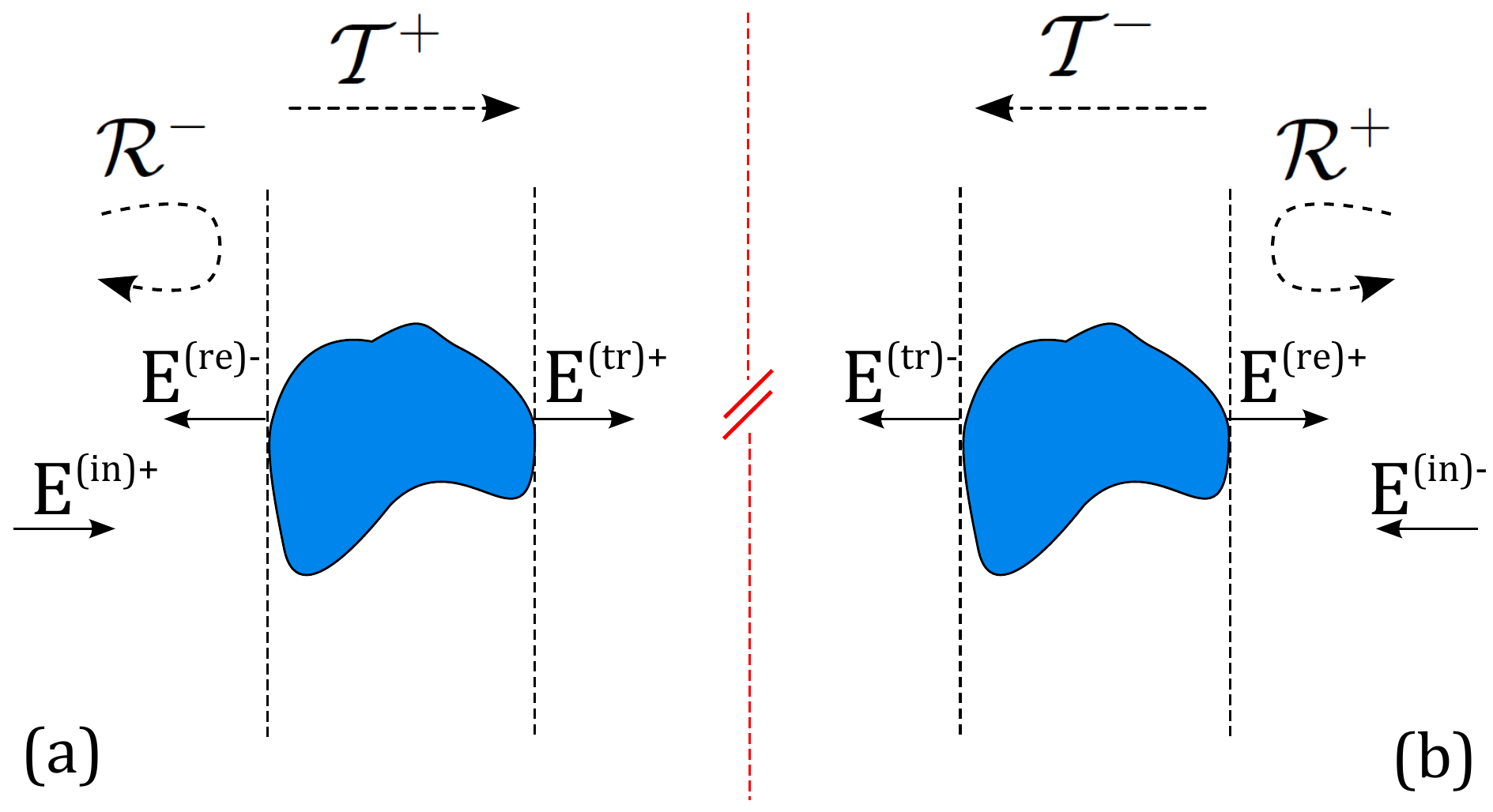}}\vspace{-10pt}
\caption{Schemes for scattering
operators.}\label{FigScattering}\end{center}\end{figure}

For a system out of thermal
equilibrium, the theorem \eqref{FluDiss} is not valid, and the
expression of the correlators is not explicit in general.
Nevertheless, in the particular case of stationary
non-equilibrium, this is possible by {\it tracing back} the
knowledge of the correlators to the description of the fields
emitted by each body alone and by the environment: this will be
done by defining the operators describing the scattering produced
by the presence of each body. To this end, taking into account a
single body (say object 1), we consider an incoming field coming
from the left side (region A in figure \ref{FigGeometry}), as
shown in figure \ref{FigScattering}(a): this field produces, in
general, a field on both sides of the body.

We call the field
produced in region A \emph{reflected} and the one in region B
\emph{transmitted}, and we introduce the operators $\mathcal{R}^-$
and $\mathcal{T}^+$ relating the amplitudes of these outgoing
fields to the amplitudes of the incoming field
$E_p^{\text{(in)+}}(\mathbf{k},\omega)$. Gathering in a vectorial
notation all the incident, reflected and transmitted modes (for
any $p$, $\mathbf{k}$ and $\omega$) $E^{\text{(in)+}}$,
$E^{\text{(re)-}}$ and $E^{\text{(tr)+}}$ respectively, the
definitions of the scattering operators read
$E^{\text{(re)-}}=\mathcal{R}^-E^{\text{(in)+}}$ and
$E^{\text{(tr)+}}=\mathcal{T}^+E^{\text{(in)+}}$. An analogous
procedure defines $\mathcal{R}^+$ and $\mathcal{T}^-$ (see figure \ref{FigScattering}(b)). If a body
is at rest, the scattering process conserves the frequency and
thus the matrix element
$\bra{p,\mathbf{k},\omega}\mathcal{S}\ket{p',\mathbf{k}',\omega'}$
of any scattering operator $\mathcal{S}$ is proportional to
$2\pi\delta(\omega-\omega')$. For convenience, we will work from
now on with scattering operators $\mathcal{S}(\omega)$ at a given
frequency $\omega$ and defined in the subspace $(p,\mathbf{k})$.
Before going further, it is useful to introduce a \emph{modified}
transmission operator $\tilde{\mathcal{T}}^\phi$ which gives only
the scattered part of the transmitted field. It is
defined by the relation
$\mathcal{T}^\phi=1+\tilde{\mathcal{T}}^\phi$ and, differently
from the ordinary $\mathcal{T}^\phi$, it goes to zero in the limit
of absence of the body.

To build the correlators, we need the expression of the
\emph{total field} in each region $\gamma=A,B,C$ of figure
\ref{FigGeometry}, which originates from the fields emitted by the
bodies and the environment. We would like to express the total
field in each region through the scattering operators. This can be
simply obtained by summing up all the possible multiple-scattering
events: for example the fields propagating in the two directions
in region B are the solutions of
\begin{equation}
\begin{cases}E^{\text{(B)}+}=E^{(1)+}+\mathcal{T}^{(1)+}E^{(3)+}+\mathcal{R}^{(1)+}E^{\text{(B)}-}\\
E^{\text{(B)}-}=E^{(2)-}+\mathcal{T}^{(2)-}E^{(3)-}+\mathcal{R}^{(2)-}E^{\text{(B)}+}\end{cases}
\end{equation}
The solutions of these equations as well as the analogous
relations for regions A and C are straightforward and will not be
given here explicitly.

We are then left with the calculation of the correlators of the
fields produced by the bodies and the environment. As far as the
environment is concerned, it corresponds to a free bosonic field
at temperature $T_3$ having correlators
\begin{equation}
\label{Co3}\begin{split}\langle E_p^{(3)\phi}&(\mathbf{k},\omega)E_{p'}^{(3)\phi'\dag}(\mathbf{k}',\omega')\rangle_\text{sym}=\delta_{\phi,\phi'}\frac{\omega}{2\epsilon_0c^2}N(\omega,T)\\
&\,\times\Rea\Bigl(\frac{1}{k_z}\Bigr)\delta_{pp'}(2\pi)^3\delta(\omega-\omega')\delta(\mathbf{k}-\mathbf{k}').\end{split}
\end{equation}
As for the bodies 1 and 2, our assumption that a local temperature
can be defined for each one, and remains constant in time,
reasonably leads to assume that the part of the total field
\emph{emitted} by each body is the same it would be if the body
was at thermal equilibrium with the environment at its own
temperature. This hypothesis, already used in
\cite{PolderPRB71,AntezzaPRL05,AntezzaPRLA06081,AntezzaPRLA06082,ObrechtPRL07,BimontePRA09},
implies that the correlators of the field \emph{emitted} by each
body can still be obtained using the fluctuation-dissipation
theorem \eqref{FluDiss} at its local temperature, where the Green
function is now associated to each body in absence of the other
one.

This procedure, together with the explicit connection between
Green function and scattering operators (directly obtained from
their definitions \cite{MessinaPrep}) allows us to obtain an
explicit expression of the correlators of the field emitted by
each body $i=1,2$: in particular, for modes propagating in the
same direction ($\phi=\phi'$) we obtain
\begin{equation}\label{Cois}\begin{split}
&\langle E_p^{(i)\phi}(\mathbf{k},\omega)E_{p'}^{(i)\phi\dag}(\mathbf{k}',\omega')\rangle_\text{sym}=\frac{\omega}{2\epsilon_0c^2}N(\omega,T_i)2\pi\delta(\omega-\omega')\\
&\times\bra{p,\mathbf{k}}\Bigl(\mathcal{P}_{-1}^\text{(pw)}-\mathcal{R}^{(i)\phi}\mathcal{P}_{-1}^\text{(pw)}\mathcal{R}^{(i)\phi\dag}-\mathcal{T}^{(i)\phi}\mathcal{P}_{-1}^\text{(pw)}\mathcal{T}^{(i)\phi\dag}\\
&+\mathcal{R}^{(i)\phi}\mathcal{P}_{-1}^\text{(ew)}-\mathcal{P}_{-1}^\text{(ew)}\mathcal{R}^{(i)\phi\dag}\Bigr)\ket{p',\mathbf{k}'}\end{split}\end{equation}
while for $\phi\neq\phi'$ we obtain
\begin{equation}\label{Coio}\begin{split}&\langle E_p^{(i)\phi}(\mathbf{k},\omega)E_{p'}^{(i)\phi'\dag}(\mathbf{k}',\omega')\rangle_\text{sym}=\frac{\omega}{2\epsilon_0c^2}N(\omega,T_i)2\pi\delta(\omega-\omega')\\
&\times\bra{p,\mathbf{k}}\Bigl(-\mathcal{R}^{(i)\phi}\mathcal{P}_{-1}^\text{(pw)}\mathcal{T}^{(i)\phi'\dag}-\mathcal{T}^{(i)\phi}\mathcal{P}_{-1}^\text{(pw)}\mathcal{R}^{(i)\phi'\dag}\\
&+\mathcal{T}^{(i)\phi}\mathcal{P}_{-1}^\text{(ew)}-\mathcal{P}_{-1}^\text{(ew)}\mathcal{T}^{(i)\phi'\dag}\Bigr)\ket{p',\mathbf{k}'}.\end{split}\end{equation}
In these expressions, which represent a crucial intermediate
result, we have introduced the notation
$\mathcal{P}_m^\text{(pw/ew)}=k_z^m\Pi^\text{(pw/ew)}$
(this definition will be also used for $m\neq1$ in the
following), where $\Pi^\text{(pw)}$ and $\Pi^\text{(ew)}$ are the
projectors on the propagative and evanescent sector, respectively.

The knowledge of the correlators \eqref{Co3}, \eqref{Cois} and
\eqref{Coio} allows the evaluation of the flux of $\langle
T_{zz}\rangle$ and $\langle S_z\rangle$ in each region. Finally,
the differences of such fluxes \eqref{Fms} provides a closed-form
analytic expression of the $z$ component of the force and the heat
transfer \emph{relative to body 1} expressed in terms of the three
temperatures $T_1$, $T_2$, and $T_3$, and of the scattering
operators of body $1$ and $2$:
\begin{equation}\label{Ffinal}\begin{split}F(T_1,T_2,T_3)&=\frac{F^{\text{(eq)}}(T_1)+F^{\text{(eq)}}(T_2)}{2}+\Delta_2(T_1,T_2,T_3)\\
H(T_1,T_2,T_3)&=\Delta_1(T_1,T_2,T_3).\\\end{split}\end{equation}
In eq. \eqref{Ffinal} we present the final result for the force as
a sum of two contributions, the first term being a thermal average
of the force $F^{\text{(eq)}}(T)$ at thermal equilibrium given by
eq. \eqref{Feq}, at the temperatures of the two bodies $T_1$ and
$T_2$ \cite{AntezzaPRLA06081,AntezzaPRLA06082}. The two terms
$\Delta_1$ and $\Delta_2$ defined in eq. \eqref{Ffinal} can be
collected as
\begin{widetext}
\begin{eqnarray}\label{Noneq}\notag\lefteqn{\hspace{-.5cm}\Delta_m(T_1,T_2,T_3)=(-1)^{m+1}\hbar\Tr\Bigl\{\omega^{2-m}\Bigl[\frac{n_{21}}{2}\Bigl(A_m(\mathcal{R}^{(2)-},\mathcal{R}^{(1)+})-(-1)^m A_m(\mathcal{R}^{(1)+},\mathcal{R}^{(2)-})\Bigr)+n_{13}\mathcal{P}_{m}^{\text{(pw)}}\mathcal{R}^{(1)-}\mathcal{P}_{-1}^\text{(pw)}\mathcal{R}^{(1)-\dag}}\\
\notag & & \hspace{-1cm} \,+(-1)^mn_{m3}\mathcal{P}_{m-1}^{\text{(pw)}}+n_{31}\Bigl[(-1)^mB_m(\mathcal{R}^{(1)+},\mathcal{R}^{(2)-},\mathcal{T}^{(1)+})-\Bigl(\mathcal{P}_{m}^{\text{(pw)}}\mathcal{R}^{(1)-}\mathcal{P}_{-1}^{\text{(pw)}}\mathcal{T}^{(1)+\dag}U^{(12)\dag}\mathcal{R}^{(2)-\dag}\mathcal{T}^{(1)-\dag}+\text{h.c.}\Bigr)\Bigr]\\
& & \hspace{-1cm} +\Bigl[n_{12}\Bigl(\mathcal{P}_{-1}^\text{(pw)}+\mathcal{R}^{(2)-}\mathcal{P}_{-1}^\text{(ew)}-\mathcal{P}_{-1}^\text{(ew)}\mathcal{R}^{(2)-\dag}-\mathcal{R}^{(2)-}\mathcal{P}_{-1}^\text{(pw)}\mathcal{R}^{(2)-\dag}\Bigr)+n_{23}\mathcal{T}^{(2)-}\mathcal{P}_{-1}^{\text{(pw)}}\mathcal{T}^{(2)-\dag}\\
\notag & & \hspace{-1cm}
+n_{13}\mathcal{R}^{(2)-}\mathcal{T}^{(1)+}\mathcal{P}_{-1}^{\text{(pw)}}\mathcal{T}^{(1)+\dag}\mathcal{R}^{(2)-\dag}\Bigr]U^{(21)\dag}\mathcal{T}^{(1)-\dag}\mathcal{P}_{m}^\text{(pw)}\mathcal{T}^{(1)-}U^{(21)}+n_{32}B_m(\mathcal{R}^{(2)-},\mathcal{R}^{(1)+},\mathcal{T}^{(2)-})\Bigr]\Bigr\}.\end{eqnarray}
Equations \eqref{Feq} and \eqref{Noneq} contain a trace, defined
by the relation
$\Tr\mathcal{A}(\omega)=\sum_p\int\frac{d^2\mathbf{k}}{(2\pi)^2}\int_0^{+\infty}\frac{d\omega}{2\pi}\bra{p,\mathbf{k}}\mathcal{A}(\omega)\ket{p,\mathbf{k}}$,
the function $n_{ij}=n(\omega,T_i)-n(\omega,T_j)$ ($i,j=1,2,3$),
and the two supplementary functions
\begin{equation}\begin{split}A_m(\mathcal{R}^{(1)+},\mathcal{R}^{(2)-})&=U^{(12)}\Bigl(\mathcal{P}_{-1}^{\text{(pw)}}-\mathcal{R}^{(1)+}\mathcal{P}_{-1}^{\text{(pw)}}\mathcal{R}^{(1)+\dag}+\mathcal{R}^{(1)+}\mathcal{P}_{-1}^{\text{(ew)}}-\mathcal{P}_{-1}^{\text{(ew)}}\mathcal{R}^{(1)+\dag}\Bigr)\\
&\,\times U^{(12)^\dag}\Bigl(\mathcal{P}_m^{\text{(pw)}}+(-1)^m\mathcal{R}^{(2)-\dag}\mathcal{P}_m^{\text{(pw)}}\mathcal{R}^{(2)-}+\mathcal{R}^{(2)-\dag}\mathcal{P}_m^{\text{(ew)}}+(-1)^m\mathcal{P}_m^{\text{(ew)}}\mathcal{R}^{(2)-}\Bigr),\\
B_m(\mathcal{R}^{(1)+},\mathcal{R}^{(2)-},\mathcal{T}^{(1)+})&=U^{(12)}\mathcal{T}^{(1)+}\mathcal{P}_{-1}^{\text{pw}}\mathcal{T}^{(1)+\dag}U^{(12)\dag}\\
&\,\times\Bigl(\mathcal{P}_m^{\text{(pw)}}+(-1)^m\mathcal{R}^{(2)-\dag}\mathcal{P}_m^{\text{(pw)}}\mathcal{R}^{(2)-}+\mathcal{R}^{(2)-\dag}\mathcal{P}_m^{\text{(ew)}}+(-1)^m\mathcal{P}_m^{\text{(ew)}}\mathcal{R}^{(2)-}\Bigr).\\\end{split}\end{equation}
\end{widetext}
The force and heat transfer on object 2 can be obtained from
\eqref{Ffinal} by changing the sign, and by interchanging indexes
1 and 2, as well as $+$ and $-$ in its explicit expression. The
term \eqref{Noneq} is purely a non-equilibrium contribution,
obeying the equality $\Delta_m(T,T,T)=0$. We remark that eq.
\eqref{Noneq} contains terms proportional to the transmission
operators $\mathcal{T}^{(1)\pm}$ and $\mathcal{T}^{(2)-}$,
resulting from taking into account the finiteness of objects 1 and
2, which were absent in previous investigations concerning
infinitely thick bodies
\cite{AntezzaPRLA06081,AntezzaPRLA06082,BimontePRA09}. It is worth
stressing that eq. \eqref{Noneq} provides a finite value of the
force and the heat transfer for any finite body 1. This property
is not evident at first sight since the expression \eqref{Noneq}
contains also divergent terms. For instance, this is the case of
the first term in the second line, proportional to the operator
$\mathcal{P}_{m-1}^{\text{(pw)}}$. Since this operator is by
definition diagonal in the $(\mathbf{k},p)$ basis, its trace is
proportional to $(2\pi)^2\delta(\mathbf{0})$, thus divergent.
Moreover, we observe that this term diverges for any choice of the
body 1, since it is formally independent on its scattering
operators. Nevertheless, the quantity \eqref{Noneq} remains finite
due to peculiar cancellations of several divergent terms. In order
to show that, we must pay attention to all the terms which, in
analogy with $\mathcal{P}_{m-1}^{\text{(pw)}}$, do not go to zero
in absence of body 1. This is the case of the operators
$U^{(12)}$, $U^{(21)}$ and $T^{(1)\phi}$ ($\phi=+,-$), having all
the identity operator as a limiting value in absence of the body
1. By making use of the definition of the
$\tilde{\mathcal{T}}^{\phi}$ operator and of the properties
$U^{(12)}=1+\mathcal{R}^{(1)+}\mathcal{R}^{(2)-}U^{(12)}$ (and
analogous for $U^{(21)}$) it is straightforward to show that all
the terms not going to zero in absence of body 1 perfectly cancel
each other, and only contributions proportional to either
$\mathcal{R}^{(1)\phi}$ or $\tilde{\mathcal{T}}^{(1)\phi}$
operators remain, leading indeed to a finite value of both the
force and the heat transfer on any body 1.

\section{Some applications}We will now apply eq. \eqref{Noneq} to two specific systems, namely atom-surface and slab-slab configurations. As anticipated before, although the derivation of the main result \eqref{Noneq} benefits from
the choice of the plane-wave basis, it can be applied to any couple of bodies (separated by a plane) even not
satisfying translational invariance. An interesting candidate to check the validity of this feature is the atom-plane system. In this case, the scattering operators associted to the atom can be deduced
by describing the atom as an induced dipole $\mathbf{d}(\omega)=\alpha(\omega)\mathbf{E}(\mathbf{R}_A,\omega)$ proportional to the $\omega$ component of the electric field calculated at the atomic position
$\mathbf{R}_A=(\mathbf{r}_A,z_A)$ through the atomic dynamical polarizability $\alpha(\omega)$ (isotropy for the atomic polarizability has been assumed). Writing the field produced by the induced dipole and projecting it on the plane wave basis as done in \cite{MessinaPRA09}
we obtain the following expression for the reflection and transmission atomic operators (for $\phi=+,-$)
\begin{equation}\begin{split}\langle\mathbf{k},p|&\mathcal{R}_A^\phi(\omega)|\mathbf{k}',p'\rangle=\frac{i\omega^2\alpha(\omega)}{2\epsilon_0c^2k_z}\Bigl(\hat{\bbm[\epsilon]}_p^{\phi}(\mathbf{k},\omega)\cdot\hat{\bbm[\epsilon]}_{p'}^{-\phi}(\mathbf{k}',\omega)\Bigr)\\
&\,\times\exp[i(\mathbf{k}'-\mathbf{k})\cdot\mathbf{r}_A]\exp[-i\phi(k_z+k'_z)z_A]\\
\langle\mathbf{k},p|&\tilde{\mathcal{T}}_A^\phi(\omega)|\mathbf{k}',p'\rangle=\frac{i\omega^2\alpha(\omega)}{2\epsilon_0c^2k_z}\Bigl(\hat{\bbm[\epsilon]}_p^{\phi}(\mathbf{k},\omega)\cdot\hat{\bbm[\epsilon]}_{p'}^{\phi}(\mathbf{k}',\omega)\Bigr)\\
&\,\times\exp[i(\mathbf{k}'-\mathbf{k})\cdot\mathbf{r}_A]\exp[-i\phi(k_z-k'_z)z_A].\end{split}\end{equation}
These expressions clearly show that the wavevector $\mathbf{k}$
and the polarization $p$ are not conserved in the atomic
scattering process, as a consequence of the absence of
translational invariance. These operators can be inserted in eqs.
\eqref{Feq} and \eqref{Noneq} in order to obtain the equilibrium
and non-equilibrium contributions to the force: to this aim, only
the terms up to the first order in $\mathcal{R}$ and
$\tilde{\mathcal{T}}$ are kept, coherently with a
second-order perturbative approach with respect to the electric
charge (first order with respect to atomic polarizability). As
shown for example in \cite{MessinaPRA09}, this gives back the
known result for the force at thermal equilibrium. As far as the
non-equilibrium contribution is concerned, taking the atom on the
left ($z_A<0$) of the surface $z=0$, a simple calculation yields
\begin{equation}\label{Atom}\begin{split}&\Delta_2(T_1,T_2,T_3)=\frac{\hbar}{4\pi^2\epsilon_0c^2}\Ima\Bigl\{\sum_p\int_0^{+\infty}d\omega\,\omega^2\alpha(\omega)\\
&\times\Bigl[n_{23}\int_0^{\frac{\omega}{c}}dk\,k\bigl(|\rho_p|^2+|\tau_p|^2-1\bigr)\\
&\,+\int_0^{\frac{\omega}{c}}dk\,k\bigl(\bbm[\hat{\epsilon}]_p^+\cdot\bbm[\hat{\epsilon}]_p^-\bigr)\bigl(n_{32}\rho_p\,e^{-2ik_zz_A}+n_{13}\rho_p^*e^{2ik_zz_A}\bigr)\\
&\,+n_{12}\int_{\frac{\omega}{c}}^{+\infty}dk\,k\bigl(\bbm[\hat{\epsilon}]_p^+\cdot\bbm[\hat{\epsilon}]_p^-\bigr)\rho_p^*
e^{-2ik_zz_A}\Bigr]\Bigr\}\end{split}\end{equation} where
$\rho_p$ and $\tau_p$ are respectively the
reflection and transmission Fresnel coefficients associated to a
planar slab and the dependence of all the quantities inside the
integral on $\omega$ and $k$ is implicit. We remark the
fact that in eq. \eqref{Atom}, the polarizability $\alpha(\omega)$
is for an atom at temperature $T_1$. The first term in the square
bracket in eq. \eqref{Atom} is a distance-independent contribution
already discussed in \cite{HenkelJOptA02,AntezzaPRL05}.
The second and the third terms are both distance-dependent: the
former depends on the propagative sector, whereas for the latter
only evanescent waves contribute. We have verified that, taking a
groun-state atom ($T_1=0$) and assuming that $T_2$ and $T_3$ are
such that atomic excitation can be excluded (which amounts to
replace $\alpha(\omega)$ by its static value $\alpha(0)$), we
recover the result of \cite{AntezzaPRL05}, obtained using a
different approach.

\begin{figure}\begin{center}\scalebox{0.48}{\includegraphics{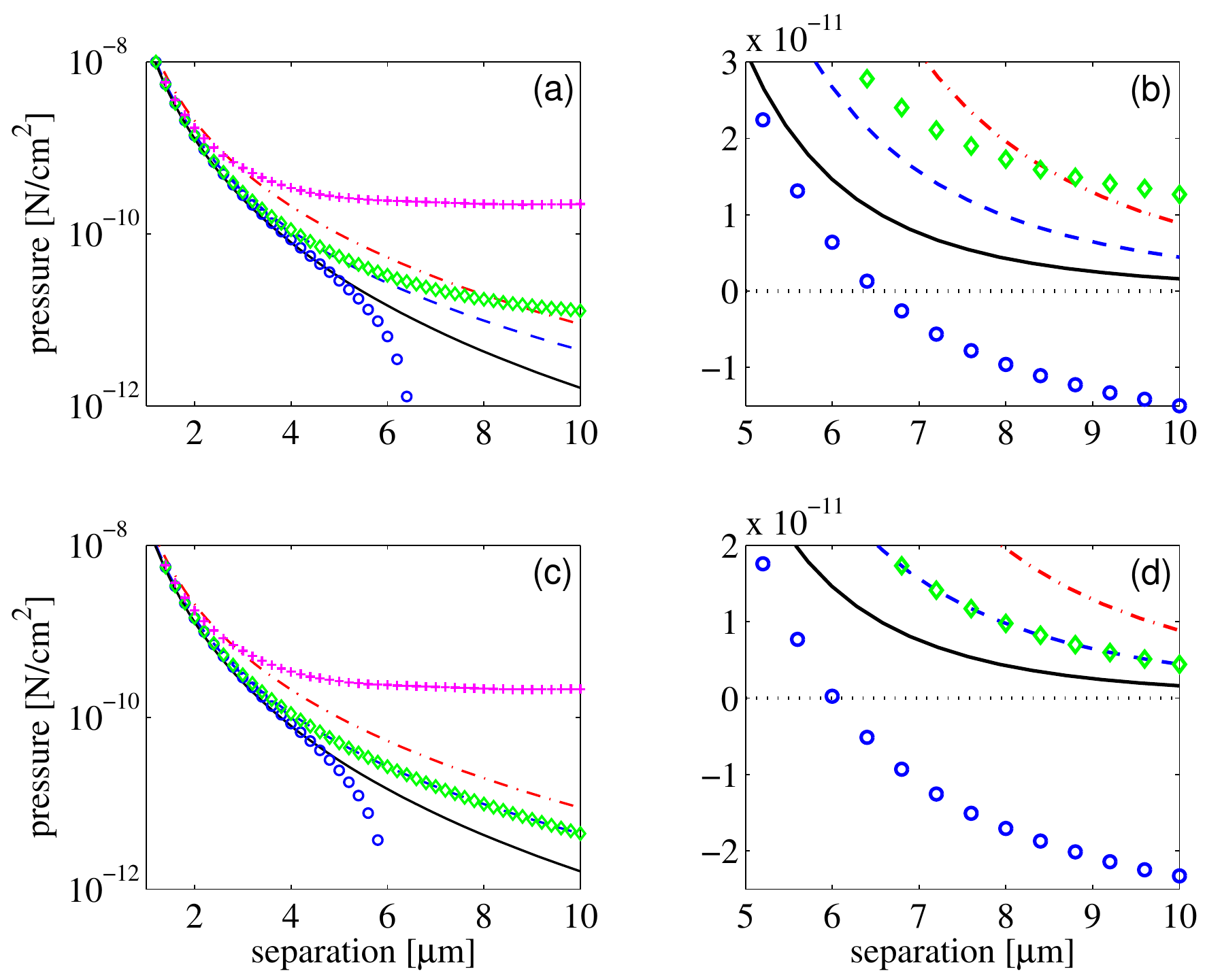}}\vspace{-10pt}
\caption{(color online) Pressure acting on a 2$\mu$m  thick slab
parallel to a 1000$\mu$m thick slab (see text). Lines: equilibrium
pressures at $T=0\,$K (solid), 300\,K (dashed), 600\,K
(dash-dotted). Symbols: non-equilibrium pressures, $T_3=0\,$K
(circles), 300\,K (diamonds), 600\,K (crosses), with $T_1=300\,$K
and $T_2=0\,$K in (a)-(b) and $T_1=T_2=300\,$K in
(c)-(d).}\label{FigForce}\end{center}\end{figure}

We consider now a second example for which we show quantitatively
the new physical features produced by these terms. We calculate
the non-equilibrium pressure on a $2\,\mu$m thick slab (body 1,
fused silica) parallel to a $1000\,\mu$m thick slab (body 2,
silicon). In figure \ref{FigForce}(a)-(b) we show the case
$T_1=300\,$K and $T_2=0\,$K, whereas in \ref{FigForce}(c)-(d)
$T_1=T_2=300$\,K: in both cases $T_3$ takes the three
values 0\,K, 300\,K and 600\,K. The figure shows that in both
cases (a)-(b) and (c)-(d) the non-equilibrium pressure can
significantly differ from the equilibrium counterpart at any of
the three temperatures involved. Moreover, both for equal and
unequal $T_1$ and $T_2$, the choice $T_3=0$ produces a repulsive
force starting around $6\,\mu$m of distance between the plates.
This is particularly remarkable in the case $T_1=T_2=300\,$K,
showing that the environmental temperature may play an important
role for objects of finite thickness, qualitatively modifying the
behavior of the force.

\section{Conclusions}We have derived a general expression for the Casimir-Lifshitz force and for the radiative heat transfer for
systems out of thermal equilibrium, valid for bodies having
arbitrary shape and dielectric function. Depending on the bodies
and on the environmental temperatures, the force and heat transfer
present several interesting degree of freedom. Due to its
generality, our results allow a straightforward  study of the
force and heat transfer for systems involving bodies whose
scattering matrices are analytically known (atoms, cylinders,
spheres and slabs)
\cite{Krugerarxiv,RodriguezArXiv11,MessinaPrep}, and
also the investigation of most general bodies by a numerical
evaluation of the scattering matrix. In particular, the heat
transfer expression will allow to obtain more accurate estimations
useful for past and future experiments as well as for technological applications such as solar cells. The force, which we
calculated explicitly for an atom in front of a slab and
numerically for two parallel slabs, can be significantly affected
by thermal non-equilibrium with the environment, presenting
transition from attractive to repulsive behaviors at distances of
few microns.

\end{document}